\pdfoutput=1

\documentclass[preprint,12pt]{elsarticle}

\usepackage{amssymb}

\usepackage{color}
\usepackage{setspace}

\begin{document}

\begin{frontmatter}



\title{A sequential algorithm with a built in tension-propagation mechanism for modeling the  chain-like bodies dynamics}

\author{Andrzej Z. Grzybowski\corref{cor1} }
\cortext[cor1]{andrzej.grzybowski@im.pcz.pl}
\author{Zbigniew Doma\'nski\corref{cor2} }
\cortext[cor2]{zbigniew.domanski@im.pcz.pl}

\address{Institute of Mathematics, Czestochowa University of Technology, PL-42201 Czestochowa, Poland}

\begin{abstract}
In the paper a novel stochastic algorithm designed to study  of chain-like bodies dynamics is introduced. This algorithm models chain movements induced by the tension propagation and its main idea relies on the sequentialization of each movement  into a sequence of virtual steps made by chain's segments. In this spirit, any accepted chain's new position is achieved by a move that is initiated by a shift of one segment picked randomly according to a problem-specific probability distribution and then followed by a cascade of some other segments' position rearrangements. The rearrangement  process  terminates when the tension in the chain induced by the initial shift is released. A considerable gain in the volume of allocated memory is achieved because the virtual steps lead to new conformations that are very likely to be acceptable by nature. We validate the algorithm by comparing passage times for polymer translocation through a pore obtained within this algorithm with their counterparts reported in the literature. In this paper we focus on a fluctuating-bond model of self-avoiding polymers on 2D square lattice. Based on the large data sets received in our simulations we have found that the transolaction time is distributed according to the Moyal probability distribution. This novel finding enables us to identify the theoretical form of various distributions of translocation time reported in literature by expressing them very accurately with the help  of this two-parameter family of probability distributions.  
\end{abstract}


\end{frontmatter}



\section{Introduction}
\label{intro}
Algorithms enable one to connect ideas and their prospective results in an efficient way. How efficient a given algorithm is depends on many factors but from the hardware perspective its speed and the volume of allocated memory are the most important. Only these two factors show that a user who is looking for a tool to solve a computationally demanding problem has to carefully choose among accessible algorithms. It is perfectly valid in the context of studies involving a chain-like assembly of prototypical beads representing an ordered in space arrangement of unit blocks.

Chain-like bodies play an important role in modeling systems of a different nature of ingredients with different types of interactions among them. Examples are an electrically charged assembly of nanoparticles,  chain-like bodies of magnetic moments with different ordering, chain structures in polydisperse ferrofluinds~\cite{ferrochains} or primary structure of proteins.  Perhaps the most known bodies of such kind are the macromolecules~\cite{biomolecules} and especially polymers which attract much attention recently \cite{multiAdMC, Markovian}. If a rigorous description of chemical details of the polymer is not necessary to capture the essential features of the biomolecule a widely employed approach relies on a coarse-grained model of a polymer chain. Such a model's chain comprises a number of substituents called beads being images of the complicated chemical monomers lying around the backbone of the real macromolecule.

Before we enumerate steps of our algorithm we would like to point to some physical aspects that the algorithm have to recognize. Computational schemes used in the field of polymer dynamics generally fall into two classes: (i) a Molecular Dynamics simulation when an appropriate second order differential equation of motion is written down, such as the Langevin equation~\cite{Langevin1, Luo} and then this equation is integrated in time by some methods~\cite{Ermak} or (ii) a Monte Carlo scenario with rules for the stochastic generation of a sequence of configuration is used to trace the system trajectory in the configuration space. Within these two classes of methods, the system is studied with regard to some constraints imposed either by a computational scheme alone or these coming directly from the field of study.  Apart from the well-established requirements such as {\it e.g.} ergodicity or self-avoidance, a modern algorithm deserved to polymer science has to also take into account the presence of the tension-propagation mechanism reflecting the response of a polymer to local drag forces~\cite{tension1,tension2,tension3}. Thus any algorithm that serves to study dynamics of a polymer should contain an inherent computational pathway ensuring such a mechanism.

Our paper is motivated by the conspicuous absence, within known algorithms, of a procedure which with high acceptance probability yields globally  deformed conformations with neither the self-avoidance nor the ergodicity violations but supporting the tension mechanism. 

In the next section the algorithm for modeling the tension propagation is introduced  in a rather formal way. Section 3 presents our results concerning the standard fluctuating-bond phantom model of self-avoiding polymer on a 2D square lattice. In these studies we examine the distribution of the passage times for polymer translocation through a pore in a membrane. Next, in Section 4,  we validate our algorithm by comparing the distribution of the passage time received in our simulation  experiments with their counterparts reported in literature. Here, also the Moyal probability distribution \cite{moyalwork} is proposed for modeling  the passage times in the considered phenomena. The article ends with some conclusions concerning the benefits resulting from the application of the algorithm as well as its possible applications.

\section{Algorithm description}
\label{algo}
Inspired by the notion of a strategy defined within the theory of  games we present an algorithm designed to study a chain-like body (CLB) propagation through an ensemble of its conformations.
In the game theory the idea of a strategy makes it possible to reduce  any sequence of possible decisions made by a player during the course of the game  to exactly one choice of {\it a strategy}, see e.g. \cite{Morris}. What is important in the  definition of a strategy is the observation that when each decision is made (in the sequence) then some {\it a priori} possible situations become no longer accessible. Consequently, when a given strategy is adopted a player does not have to  consider all {\it a priori} possible game "configurations" but only those which can be accessed with the help of this strategy. We adopt the same idea in our algorithm. From a given instant conformation the chain body skips to its another allowable conformation according to a strategy adopted by Nature.  Within this algorithm a move between two consecutive conformations is built up from a set of virtual steps related to elements of the chain. 

In this section we introduce the algorithm in a formal fashion. For this purpose we need to introduce some terminology.

An abstract $2D$ CLB {\it position} is a finite sequence $\mathbf{c}=\left\{ c_1,c_2,\ldots,c_n\right\}$ of $2D$ points $c_i=\left(x_i,y_i\right)$ such that the distance $d$ between any pair of its successive elements is bounded by given limits $L_{min}, L_{max}$, i.e. $L_{min}\leq d(c_i,c_{i+1})\leq L_{max}$. 
 The limit $L_{max}$ reflects some constraints as e.g., constraints imposed by interactions between segments. For example, the Finite Extensible Non-Linear Elastic (FENE) potential includes the maximum bond extension, i.e. the maximal allowed separation between two consecutive monomers equals to $L_{max}$. 

The elements of the sequence $\mathbf{c}$ are called {\it segments' positions} (shortly {\it segments}) of the CLB. 

{\bf Assumption 1} (discretization of the motion space): The CLB moves along the integer lattice nodes, i.e. coordinates of each segment $\left(x_i,y_i\right)$ are integer.

It is common in the literature to distinguish between studies conducted within continuous description of polymer-like bodies and these mapped onto a lattice with a coarse grained view on the CLB structure. Both sorts of description are related by some relevant functional-integral limits and the resulting continuous models yield similar characteristics as the corresponding discrete models. Discrete models themselves may depend on continuous space variables or they are formulated with use of variables defined on the countable sets of points, as e.g. on a lattice nodes.  However, even in the discrete version to achieve better approximation of the continuous motion space,  one may assume that the abstract unit of the length (distance) is equivalent to  a given number of grid sides. Obviously, the greater is the number of grid sides per unit  (GSPU), the better approximation of the continuous space. On the other hand greater values of GSPU lead to more computationally demanding simulations.

Any pair $\left(c_i,c_{i+1}\right)$ of consecutive segments is called an ($i-$th) {\it edge}, $i=1,2,\ldots,n-1$. Number $n$ is called the {\it length} of the CLB. The {\it structure} of the CLB is defined by the relative mutual positions  and orientations of all segments and edges. 

In the description of the algorithm we distinguish {\it steps} and {\it moves}.  A {\it step} is made by a single segment of the CLB,  carrying it from one lattice node to another. 
The steps made by segments may be influenced by various external {\it laws}. For example there may exist a variety of driving forces generated externally. Such outer rules may favor certain lengths or directions of a step. 
In our algorithm the impact of such external laws  can be  incorporated by a specific probability distribution $\pi$ defined on a set of nodes that can be reached by a segment in a single step. This set will be called {\it one step reachable nodes} (OSRN) and consists of nodes (say $b$), which for a given segment $c$  satisfy the following condition:
$d\left(c,b\right)\leq R_{max}$. 
The parameter $ R_{max}$ equals assumed maximal length of a step made by a segment in a case where   no other (internal) forces and/or restrictions are present. This probability distribution determines the more and the less probable directions and/or lengths of the steps. Such a {\it one step probability distribution} (OS p.d.) may also depend on the segment's coordinates (i.e. its position in a motion space).

Due to assumed properties of the environment and as a result of assumed features of the CLB itself, for any given segment among its OSRN there may exist {\it actually forbidden nodes} (AFN).  For example, one of such restrictions is the upper  limit for the distance between subsequent   segments. Such a restriction assures the continuity of the CLB. Another example of this kind of restrictions may be a requirement that in a given node, at most a given number of segments can be placed (e.g. the repton model \cite{Rub87, repton}, self-avoidance restriction etc). Yet another example is the existence of different objects that already occupy some nodes, such as a cell's membrane or any other kind of obstacles~\cite{lattice-obstacles}.

For any given segment the subtraction  of the sets OSRN and AFN will be called a set of {\it actually accessible nodes} (AAN): $\mbox{AAN}=\mbox{OSRN}\setminus \mbox{AFN}$. The OS p.d. truncated to the set AAN will be called {\it actual step probability distribution} (AS p.d.)

Let us illustrate the introduced notions in exemplary graphs.
In these graphs  we assume $\mbox{GSPU}=2$ and the following values of remining parameters: $L_{min}=1$,  $L_{max}=3$ and $R_{max}=4$ (i.e. correspondingly 2, 6 and 8 grid sides). 

\begin{figure}[h]
\begin{center}
\begin{minipage}[b]{6.2cm}
\centering
\includegraphics[width=6cm]{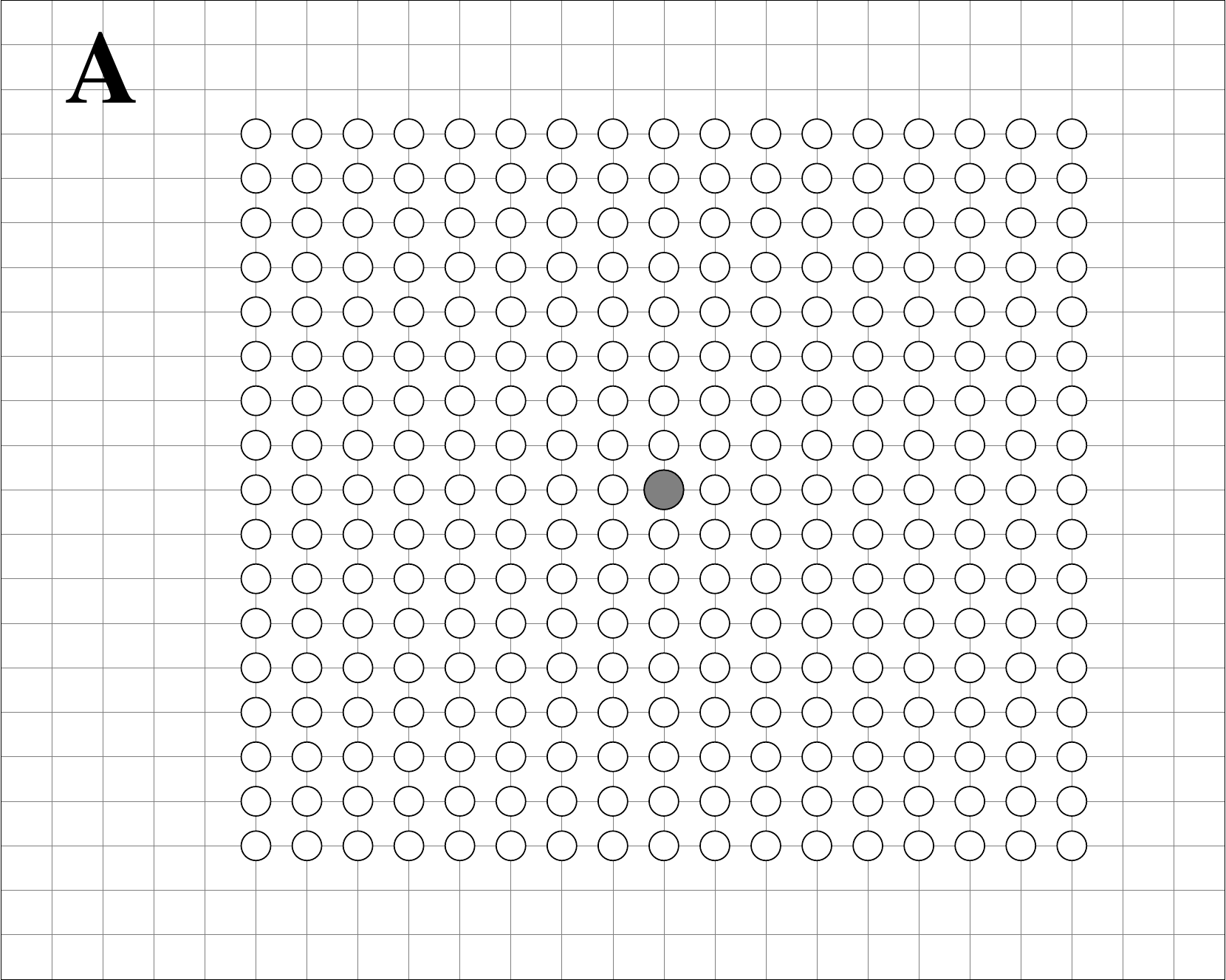}
\end{minipage}
\begin{minipage}[b]{6.2cm}
\centering
\includegraphics[width=6cm]{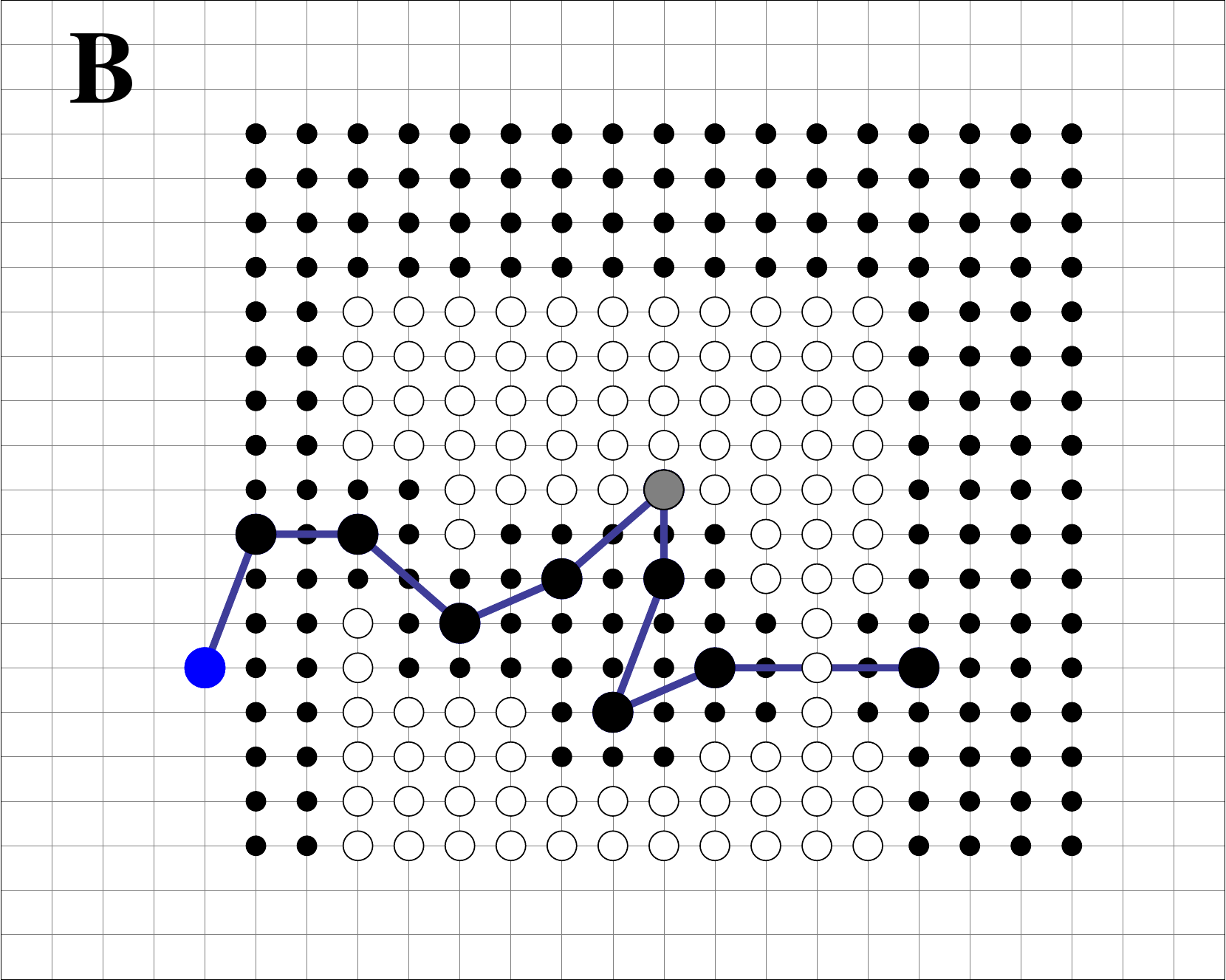}
\end{minipage}
\caption{Exemplary segment (indicated by the gray bullet) and its set of OSRN ( empty circles in graph A). Sets of AFN (black dots and bullets) and AAN (empty circles) related to the  indicated by gray bullet segment are presented in graph B. 
 The assumed values of the problem parameters: $L_{min}=1$,  $L_{max}=3$, $R_{max}=4$ and $\mbox{GSPU}=2$ }
\label{figAAN}
\end{center}
\end{figure}

The algorithm does not assume any specific  form of the distance function $d(\cdot,\cdot)$. It can be implemented with any metric which is suitable for a description of a given physical process. In our examples  the distance between segments $c=(x,y)$ and $c'=(x',y')$ is defined by a metric:
\begin{equation}\label{OurMetric}
d(c,c')=\mbox{max}\left\{\mid x'-x\mid,\mid y'-y\mid \right\}
\end{equation}
In the topology induced by the metric (\ref{OurMetric}) any sphere with the center $c$  and radius $r$ (in terms of grid sides) consists of all nodes lying in a square with the same center and with sides parallel to the axes having a length $2r+1$ . Under our assumptions about the parameters' values, the set OSRN connected with the segment $c$ indicated by the gray bullet in Fig. \ref{figAAN}A is the union of all nodes indicated by the empty circles. Consequently, the OS p.d. should be defined on this OSRN set. 
If all directions and lengths of steps are equally possible, then the OS p.d. is the uniform one.  For simplicity of the illustration let us assume the latter case in our example.

To illustrate the notions of the sets AFN and AAN let us assume that the segment  $c$ (gray bullet) belongs to the CLB that is indicated by one blue  and eight black bullets connected by the blue edges in Fig. \ref{figAAN}B. In our case the only restrictions result from the values of parameters connected with the assumed elasticity properties of the CLB i.e. the minimal and maximal distance between successive segments (no other bodies occupy the neighbouring  nodes). Thus the set AFN consists of the nodes indicated by black dots and bullets while the set of AAN is indicated by the empty circles.

By making a step towards its new position $c'_i$ the segment  $c_i$ may {\it create a tension} in the CLB structure. It is assumed that the tension is big enough to drag neighbouring segment if  $d\left(c_{i-1},c^{'}_i\right)> T_d$ and/or $d\left(c^{'}_i,c_{i+1}\right)> T_d $, where $T_d$ is a tension parameter.
A {\it move} of the CLB is a sequence of consecutive  segments' steps, that finally leads to the relaxation of the structure's tension created by the initial step.  In other words, the move reflects the tension propagation and it  is finished as soon as there is no tension in the structure. Obviously sometimes the move may consist of a single step only.

{\bf Assumption 2} (sequentialization of the CLB move): Every move of the CLB is initialized by only one segment. Then each move of the CLB can be sequentialized into a sequence of {\it steps}.

The segment that  initializes a move of the CLB will be called a {\it first to step } (FTS) segment. In the algorithm the choice of the FTS segment is random and realized according a given probability distribution defined on all CLB segments. The distribution will be denoted as FTSC p.d. The FTSC p.d. may model various physical aspects of the CLB, e.g. the constrained motion of a tagged monomer or forced relocation of a polymer capped with specific end monomers.

\begin{figure}[h]
\begin{center}
\begin{minipage}[b]{6.2cm}
\centering
\includegraphics[width=6cm]{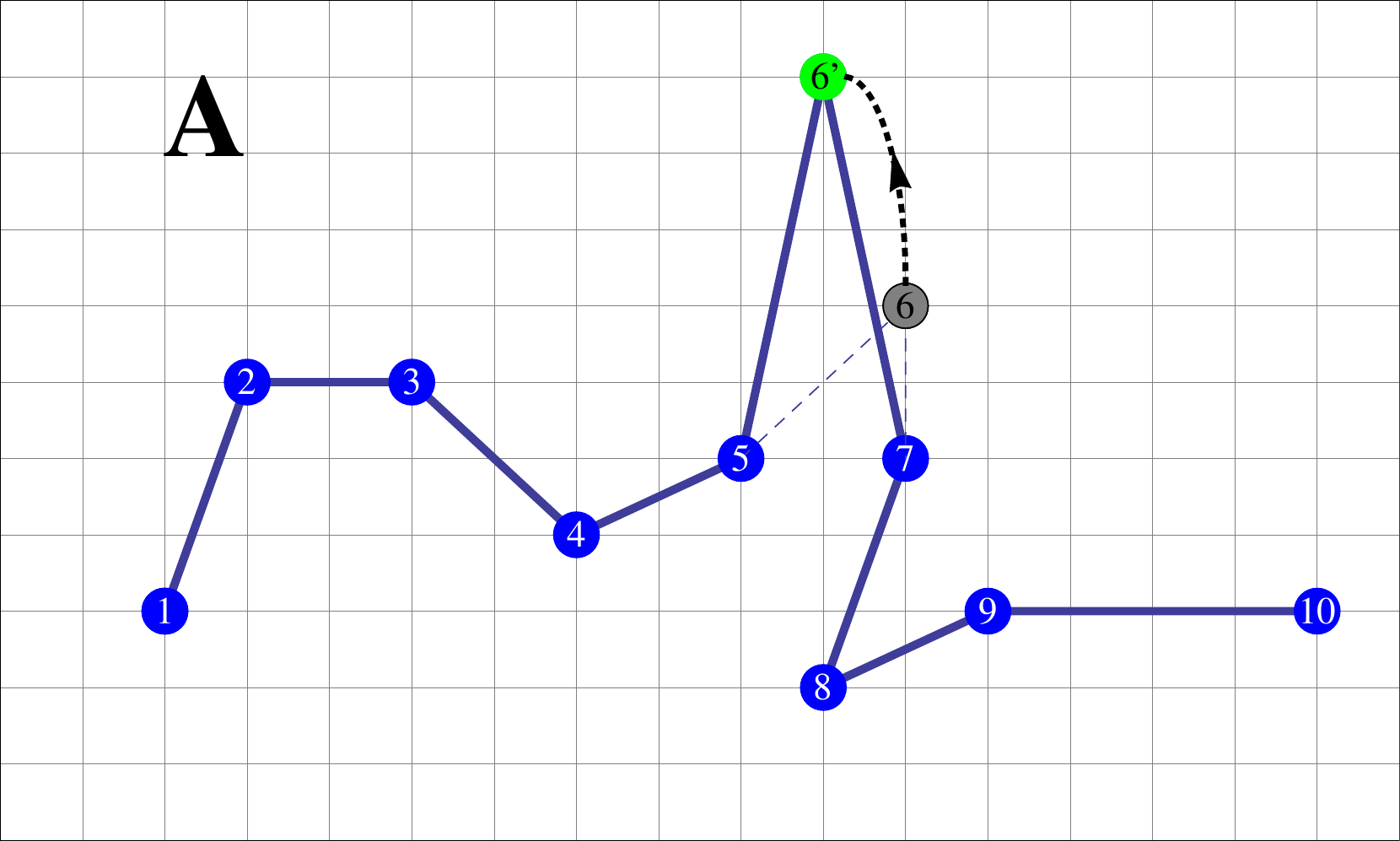}
\end{minipage}
\begin{minipage}[b]{6.2cm}
\centering
\includegraphics[width=6cm]{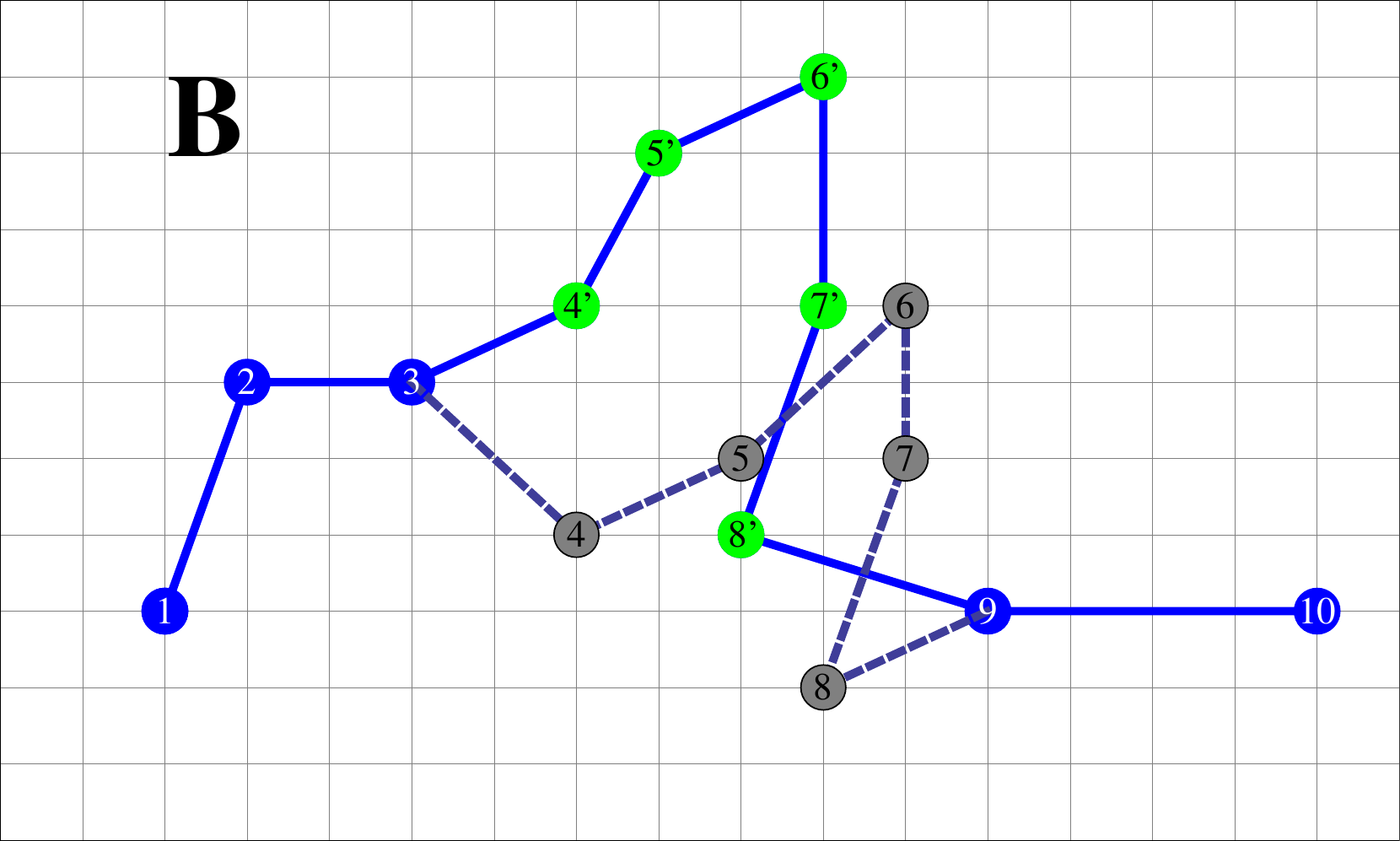}
\end{minipage}
\caption{Exemplary move made by a CLB.  
The assumed values of the problem parameters: $L_{min}=1$,  $L_{max}=3$, $R_{max}=4$ , $T_{d}=2$ and $\mbox{GSPU}=2$ }
\label{figMove}
\end{center}
\end{figure}

An example of a move made by the CLB from Fig. \ref{figAAN}B  is illustrated  in  Fig. \ref{figMove}. First let us number all segments: $c_1...c_{10}$. In  Fig. \ref{figMove} the letter "c" is omitted for clarity of the graphs. Let us also assume  that the tension parameter is $T_p=4$ and let us consider the case where the segment $c_6$ was carried from its initial position to the new  $c'_6$ indicated by the green bullet (the letter "c": is again omitted). Now  the distances between the new position $c'_6$ of $c_6$ and its neigbouring segments $c_5$ and $c_7$ are greater than $T_p$. Thus both latter segments have to make the steps. If their new positions $c'_5$ and $c'_7$ are as indicated by the appropriate green bullets in Fig.  \ref{figMove}B then again the tension is created between the pairs of segments $c_4,c'_5$ and $c'_7,c_8$. Consequently, the segments $c_4$ and $c_8$ have to also make their steps. However, because their new positions  $c'_4$ and $c'_8$ do not create any tension in the CLB, the move is completed. The final new position of the CLB is indicated by blue and green bullets in  graph B in Fig. \ref{figMove}. Note that during the run of the simulation all  new positions of the segments are chosen according the related AS p.d.

A {\it movement trajectory} is a sequence of consecutive CLB positions stored in matrix $\mathbf{C}$ whose $i$-th row is interpreted as a CLB position after $i-1$ moves (at moment $i$). Thus element $C_{ij}$ denotes the  segment $j$ in the CLB position at the moment $i$.

In many real-world situations, such as biopolymer behaviour inside a living tissue, one should also take into account some additional constraints connected with the biochemical nature of the system. Thus we define additionally the cost connected with the CLB structure. The cost of the CLB structure and its location in the motion space is a function $F$ representing its {\it fitness} connected with its conformation and/or other external (e.g. environmental) properties. The lower the cost, the better the fitness of the polymer structure and position.

{\bf Assumption 3} (acceptance of new CLB position): The new position of the CLB is accepted with a probability depending on its cost.

The above three assumptions and ideas are implemented in the following sequential algorithm for the CLB movement simulation:

{\bf Step 0}. ({\it Initialization}) Set the initial (current) CLB position $\mathbf{c}_{curr}$  and evaluate its {\it current} cost function value $F_{curr}$.

{\bf Step 1}. ({\it FTM segment selection and computation of its AS p.d.}) According the given FTMCD, select FTM segment $c_{curr,f}, f\in \left\{1,\ldots,n\right\}$. For the given segment $c_{curr}$ {\bf determine} sets OSRN, AFN and AAN as well as the AS p.d..

{\bf Step 2}. ({\it Step choice for FTM segment)} According to computed AS p.d. randomly select one node out of the set AAN for the next position of the segment $c_{curr}$.

{\bf Step 3}. ({\it Successive steps of remaining segments)} To obtain a new CLB position $\mathbf{c}_{new}$ sequentially choose segments $c_{new,i},i=f-1,f-2,\ldots$, {\bf determine} their sets SRN, AFN, and AAN as well as the ASPD. Then according to computed AS p.d. randomly draw their subsequent new positions $c_{new,i}$. This process is terminated for the first $k$,$f-1\geq k\geq1$ for which the following condition holds:$d\left(c_{curr,k},c_{new,k+1}\right)\leq T_{d}$. If $k>1$ then $c_{new,i}=c_{curr,i}$ for $i=1,\ldots k$. Next sequentially choose segments $c_{new,i},i=f+1,f+2,\ldots$, {\bf determine} their sets SRN, AFN, and AAN as well as the related AS p.d. Then according to computed AS p.d. draw randomly their subsequent new positions $c_{new,i}$. This process is terminated for the first $k$,$f+1\leq k\leq n$ for which the following condition holds:$d\left(c_{curr,k},c_{new,k-1}\right)\leq T_{d}$. If $k<n$ then, for $i=k,\ldots n$ assume $c_{new,i}=c_{curr,i}$ .

{\bf Step 4}. ({\it Acceptance of new position)} Compute the cost of the new CLB position $F_{new}$. If $F_{new}<F_{curr}$ accept $\mathbf{c}_{new}$. If $F_{new}\geq F_{curr}$, accept $\mathbf{c}_{new}$ only if random variable $U$ having a uniform probability distribution on interval $\left[ 0,1\right]$ satisfies $U\leq \psi \left( F_{\mbox{\footnotesize{new}}}-F_{\mbox{\footnotesize{curr}}}\right)$, with $\psi$ being a given {\it nonincreasing} function. If $\mathbf{c}_{new}$ is accepted then $\mathbf{c}_{\mbox{\footnotesize{curr}}}$ is replaced by $\mathbf{c}_{\mbox{\footnotesize{new}}}$; else $\mathbf{c}_{\mbox{\footnotesize{curr}}}$ remains as is.

{\bf Step 5}. {Terminate} the algorithm if the stopping criterion is met; {otherwise} return to Step 1.

{\bf Step 6}. Return the final position of the chain, its cost and required  statistics connected with the simulated movement trajectory.

The nonincreasing function that appears in Step 4 of the above algorithm represents the attitude of nature towards the acceptance of worse states. If nature accepts all states, one may assume that $\psi \equiv 1$. Otherwise, similarly as in the famous Metropolis algorithm we propose the use of function $\psi \left(z\right)=\exp \left[-z/T\right]$, where $T>0$ is a parameter which can be additionally subject to change during the movement process.

\section{Exemplary test problem}
We have tested our algorithm with some polymer models, that were discussed in the literature and we were able to reproduce the results reported previously. Moreover, the results we have obtained with the help of our algorithm allowed us to formulate more precise hypothesis concerning some theoretical aspects of the considered phenomena.

Here, our algorithm is applied to one of the simplest type of the CLS. In our simulations of the self-avoiding polymer we analyze a fluctuaiting-bond model in 2D with the purpose to find the relationship between the times consumed by this CLS when it passes through the opening in a flat membrane and the length of the CLS. In the literature different assumptions concerning the initial position of the polymer are considered. In our experiment, we organize the CLS passage in a way that the head of the chain is placed in the pore and the remaining segments are randomly arranged.  Then, a small bias induced by the OS p.d. drives polymer segments through the pore. During the simulation we keep the following values of the algorithm parameters: $L_{min}=1,  L_{max}=3, R_{max}=4, T_d=2$ and $GSPU=2$.

Within this scenario we have collected large data sets containing information about translocation time ($TT$) and corresponding lengths of the CLS. When we tried to model the randomness of $TT$ it appeared that we were able to fit  the data received in a number of different experiments with one  family of probability distributions. Surprisingly, it is the family of the Moyal p.d. that was introduced in the field of nuclear physics~\cite{moyalwork}. 

\begin{figure}
\center
\includegraphics[width=3.8in]{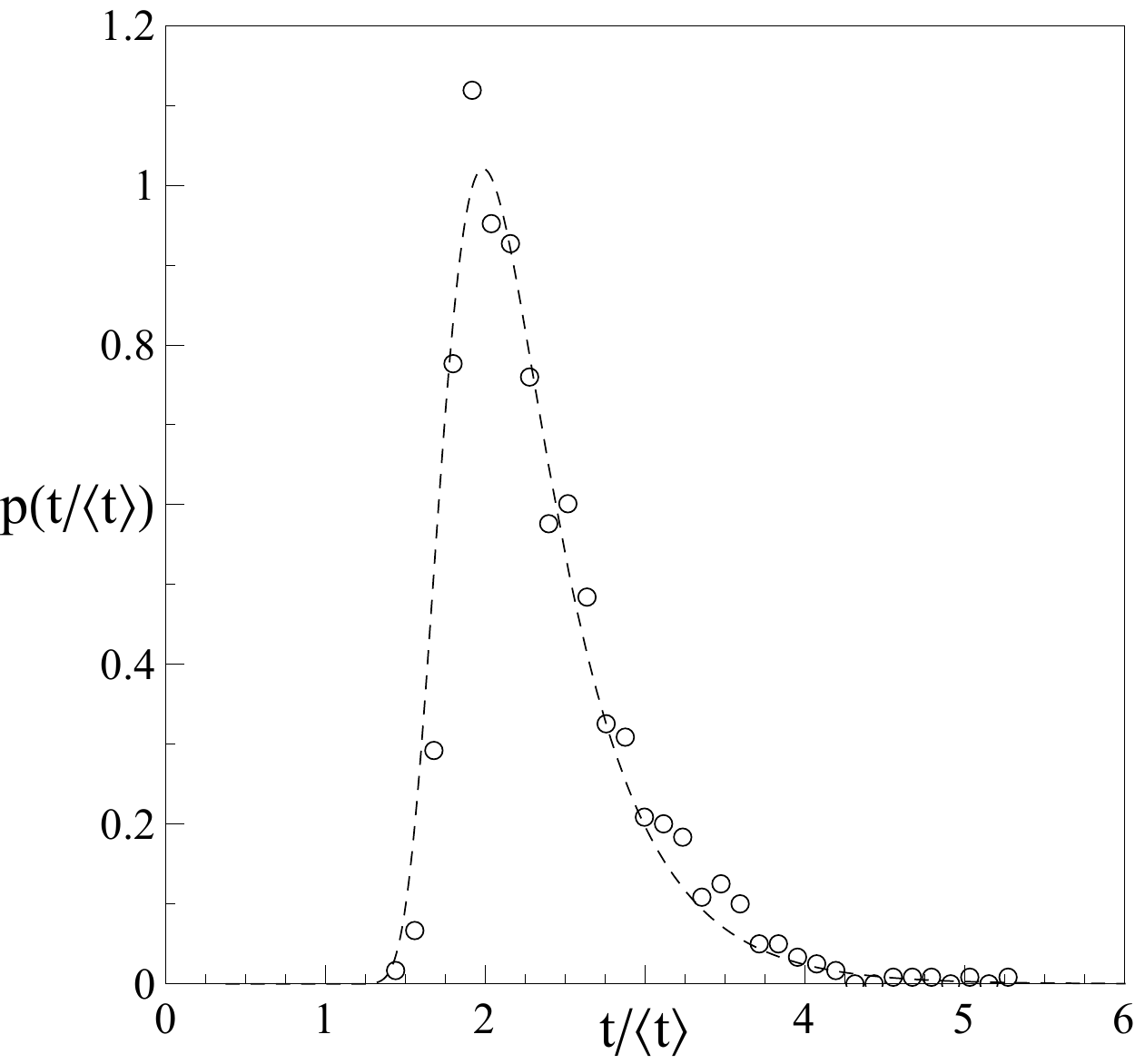}
\caption{\label{moyal-pdf}A typical translocation time distribution of polymer model computed by the presented algorithm. Here, $N=54$, the population size equals to $2\times 10^3$ and the scaling factor $<t>=N^2$ is used. The dashed line is the Moyal p.d., Eq. (\ref{phpdf}) with $a=1.989$ and $b=0.237.$}
\end{figure}

So, according to our discovery the distribution of translocation times of the $N$-segment phantom polymer is given by the following density functions
\begin{eqnarray}
p\left(a,b;N^{-2}t\right)&=&\frac{1}{b}f_{\mbox{\scriptsize{Moyal}}}\left(\frac{N^{-2}t-a}{b}\right), \label{phpdf}\\
f_{\mbox{\scriptsize{Moyal}}}(z)&=&\frac{1}{\sqrt{2\pi}}\exp\left(-\frac{z+e^{-z}}{2}\right) \label{moyal}
\end{eqnarray}
where the parameter $a$ is the peak's ordinate and the parameter $b$ scales the standard deviation $\sigma$ of the Moyal p.d., i.e. $b=(\sqrt{2}/\pi)\sigma$. Figure (\ref{moyal-pdf}) shows both a typical $TT$ distribution received on the basis of our results and the appropriate Moyal p.d. fit. 

\begin{figure}
\center
\includegraphics[width=3.6in]{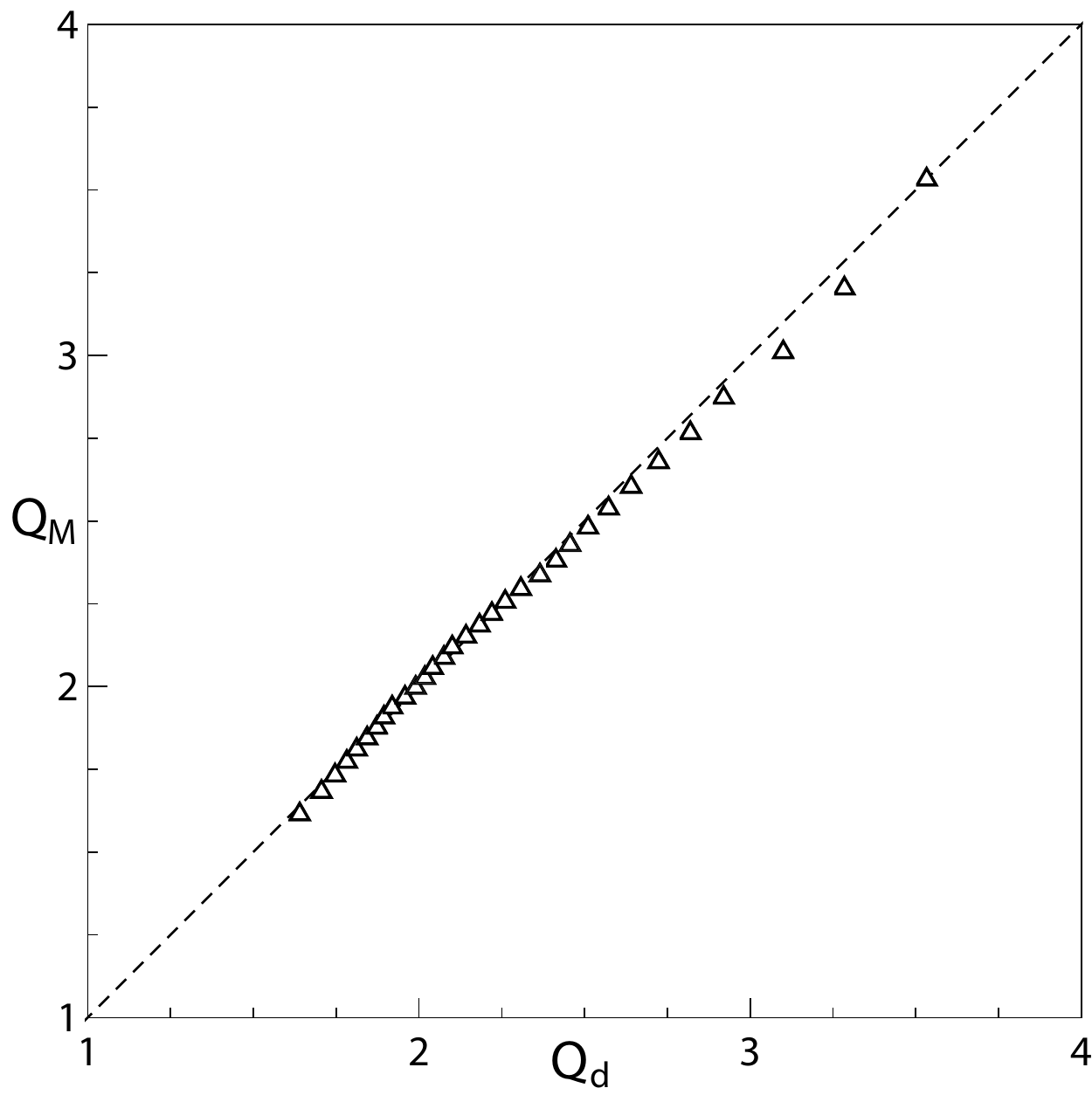}
\caption{\label{Qplot}Q-Q plot of the quantiles of the set of computed translocation time $Q_{\mbox{d}}$ vs. the quantiles $Q_{\mbox{M}}$ of the Moyal probability distribution. According to the same data as in Fig. (\ref{moyal-pdf}).}
\end{figure}

In Fig. (\ref{Qplot}) we present a quantile-quantile plot (Q-Q plot) of the quantiles of one of the collected data set against the corresponding quantiles given by the Moyal p.d. The points are well arranged along the straight line which indicates that the set of analyzed data comes from the population with the underlying Moyal distribution.

We display this Q-Q plot only for a visual-rude estimate purpose. However, we have rigorously examined the data sets using a number of goodness of fit tests. We have also  estimated values of the parameters $a$ and $b$ by employing the maximum likelihood procedure. Since our illustrative example involves only one model parameter (the number of segments $N$) we have estimated the functional dependences of the Moyal p.d. as follows

\begin{equation}
\label{abestimates}
a(N)=2.718-1.598 N^{-1/5} \hskip 9pt\mbox{and}\hskip 9pt b(N)=0.172+4.886 N^{-1}. 
\end{equation}

Using these estimates we can relate the mean $\mu$ and the standard deviation $\sigma$ with the length of CLS $N$ by the formulas

\begin{eqnarray}
\mu(N)&=&a(N)+(\ln2+\gamma)\cdot b(N) \label{average-t}\\
\sigma(N)&=&\frac{\pi}{\sqrt{2}} b(N)
\end{eqnarray}
where $\gamma\approx0.577$ is the Euler's constant. Finally, Eqs. (\ref{average-t}) and (\ref{abestimates}) relate the mean translocation time $<t>$ with $N$.

So, as we can see, the Moyal distribution perfectly well describes results received in our simulation experiments. What is even more important it also describes amazingly well results already reported in the literature. This fact    fully confirms that our algorithm is a good tool for modelling chain-like bodies dynamics. 

It is also worth emphasizing that in these above-mentioned already published results the theoretical form of the $TT$ distribution was either unknown or unsatisfactory. To the best of our knowledge, the only interesting proposal for theoretical formula for $TT$ distribution was published in \cite{Lubensky}. 
We compare some already known proposals for the analytical form of the $TT$ distribution  with the Moyal one in the next section.

\section{The Moyal probability distribution as a model for the distribution of translocation times}

In the literature of polymer science an ample set of publications deals with the distribution of time consumed by a polymer when it traverses a membrane's opening. Among these publications we have found numerous examples of data charts and/or histograms that represent results obtained in various studies, both simulation and experimental. It appears that these experimental distributions can be very accurately approximated with the appropriate Moyal p.d., see e.g.  \cite{Lubensky, Milchev, meller, huopaniemi, 2probability, cacciuto}. As we have already mentioned, first we compare our proposal with the  theoretical and experimental results presented in \cite{Lubensky, meller}.

\begin{figure}
\center
\includegraphics[width=3.6in]{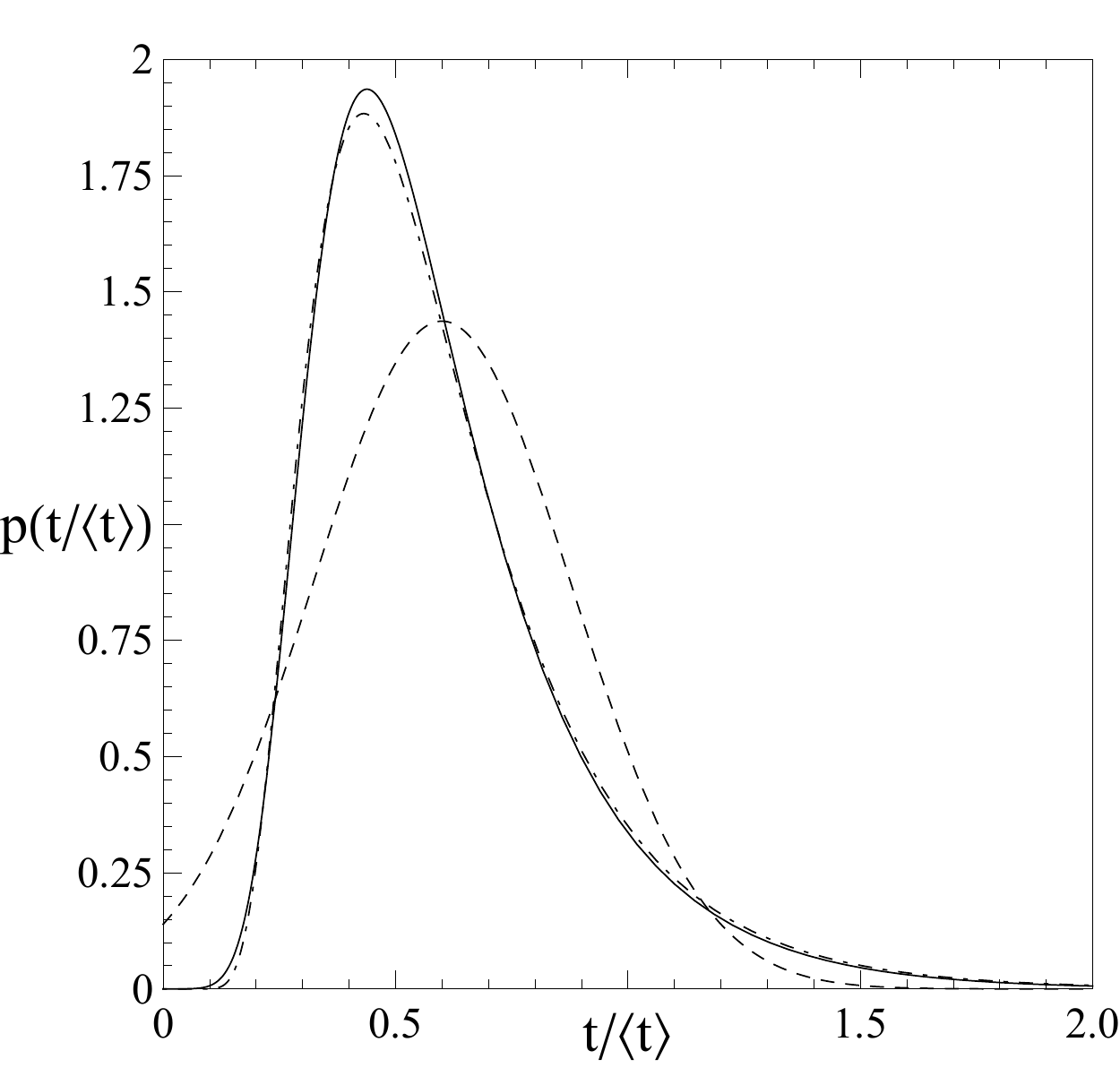}
\caption{\label{figLubensky}The probabilty density functions $p$ of translocation times vs. $t/<t>$, where $<t>$ is the mean translocation time. Solid line: the Moyal distribution, Eq. (\ref{moyal}) with $<~t~>=N^2$; dash-dotted line: the scaled distribution $\xi^{-1}\psi$, Eq. (\ref{psi}) with $<t>=L/\overline{v}=1/\xi$. The dashed curve represents a Normal distribution. All these distributions have the same means $(\mu)$ and standard deviations $(\sigma)$. Here, the values $\mu=0.6$ and $\sigma=0.2777$ yield a correspondence with the Figure 2 in \cite{Lubensky}.}
\end{figure}

Figure (\ref{figLubensky}) shows a remarkable agreement between the Moyal p.d. and the distribution $\psi$ that has been derived from the coarse-grained equation for polymer's diffusion with drift~\cite{Lubensky}. Since the distribution $\psi$, in the original paper~\cite{Lubensky}, is not normalized, for the purpose of this comparison, we have rescaled it with a factor $Z$. After that $\psi$ gets the following nondimensionalised form 

\begin{eqnarray}
\xi^{-1}\psi\left(\lambda,\xi;t\right)&=&\frac{1}{2\sqrt{\pi}Z(\lambda)\lambda^{3/2}}\cdot\frac{1-2\lambda\xi t}{(\xi t)^{7/2}\cdot\exp\left[\frac{\left(\xi t-1\right)^2}{4\lambda(\xi t)}\right]} \label{psi} \\
\xi&=&L^{-1}\overline{v} \label{ksi} \\
\lambda&=&L^{-1}l_d \label{lamb}
\end{eqnarray}
where: $L$ is the length of polymer, $l_d$ represents the diffusive length and $\overline{v}$ stands for an average polymer velocity.
As a probability distribution, the function $\psi$ must be nonnegative. Thus, for any given value of $\lambda$, the domain of $\psi$ is bounded by the condition $\xi t\leq 1/(2\lambda)$ and, in consequence, $\psi$ can not be the valid pdf for sufficiently large $t$. On the other hand, from Eq. (\ref{moyal}), it is seen that the Moyal p.d. does not suffer from such a limitation.

The function $\psi$, Eq. (\ref{psi}) has also appeared in the context of a typical histogram of voltage-driven DNA translocations through a single $\alpha$-hemolysin protein pore~\cite{meller}. A histogram, such as presented in the Fig. 3 in \cite{meller}, can easily be drawn by employing the Moyal p.d. Specifically, the width to peak ratio $\delta t_p/t_p\approx 0.55$ enables us to compute the value of the parameter $b$ in the Moyal p.d. Since in \cite{meller} $\delta t_p$ is defined at $t_p/\sqrt{e}$,  the Eqs.~(\ref{phpdf}-\ref{moyal}) yield: $\delta t_p/t_p=b^2\cdot\sqrt{2e\pi}\left[\mbox{ProductLog}(-1/e^2)-\mbox{ProductLog}(-1,-1/e^2)\right]$ and thus $b\approx 0.21$. Of course, before we wrap the histogram in the Moyal p.d. we have to pass from $t$ to $t/<t>$ or to scale the parameters $a$ and $b$  by multiplying them by $<t>$.

Apart from $\psi$, other functions were also proposed in the literature to fit distributions of data sets. One such proposal is a function $t^a_1\exp(-a_2t)$. However, this function has one serious drawback. Namely, it displays systematic deviations from the data at small $t$ (as it is seen, e.g. in \cite{Milchev}). In contrast to this drawback, the Moyal p.d. does well in handling such data. In particular, an average over $\delta t=10$ neighboring entries from the Fig. (6 b) in \cite{Milchev} is nicely fitted by the Moyal p.d. with $a=200$ (position of the peak) and $b=46$.

Another clear evidence of the Moyal p.d. presence in the description of distributions of translocation times emerges from the results of Langevin dynamics simulations of a polymer passage through a nanopore under a pulling force~\cite{huopaniemi}. The distribution presented in Fig. 2, in \cite{huopaniemi} accurately follows the Moyal p.d with the parameters $a=2.35$ and $b=0.17$.

\section{Conclusions}
\label{summary}

All of the above examples show that results received previously with the help of various techniques and approaches are closely related to these obtained with the help of our algorithm. 
It confirms the adequacy of the idea underlying our algorithm and its usefulness for modeling the CLB movements. Although in physics literature, particularly in polymer studies, the problem of  the elasticity of the bonds is  treated as very important one, such algorithms that model the tension propagation have not been reported in the stochastic simulation literature as yet. Thus the presented algorithm is especially useful in these problems where the whole CLB structure demonstrates elastic properties, and the tension propagation cannot be neglected. 

Our approach also has another advantage. The incorporated  tension propagation mechanism leads generally to new states that are more likely to be accepted by Nature what results in significant reduction of the algorithm search space. Consequently, algorithms  based on the presented idea are much faster than those that have been already presented  in literature. 
Due to increased efficiency, the new algorithm enables the researcher to  produce large data sets containing the information about the phenomenon under study. 
In our case it allowed us to identify the Moyal distribution as a very good model for the translocation time. 

The application of our algorithm also allows the researcher to take into account various additional features of the CLS and examine the relationship between a number of movement and environmental parameters. In the near future we are going to examine the mutual impact  of various movement parameters incorporated in our algorithm  (such as $ L_{min},  L_{max}, R_{max} $ or $T_d$ ) on the distribution of the trenslocation time of the CLB. 

The final conclusion is that the proposed algorithm promises affordable and robust method to study polymer movements.
\newline
\newline
\newline
{\bf References}

\biboptions{compress,square}

\end{document}